\documentclass[preprint]{aastex}

\usepackage{graphicx,natbib}

\shorttitle{GRB Spectral Lags}
\shortauthors{L. Chen, Y.-Q. Lou,et
al.}

\begin{document}

\title{Distribution of Spectral Lags in Gamma Ray Bursts}

\author{Li Chen\altaffilmark{1}}
\affil{Department of Astronomy, Beijing Normal University, Beijing
100875, P.R.China} \email{chenli@bnu.edu.cn}
\author{Yu-Qing Lou\altaffilmark{2}}
\affil{(a) Physics
Department and the Tsinghua Center for Astrophysics (THCA),
Tsinghua University, Beijing 100084, China; (b) Department of
Astronomy and Astrophysics, The University of Chicago, 5640 South
Ellis Avenue, Chicago, IL 60637 USA; (c) National Astronomical
Observatories, Chinese Academy of Sciences, A20, Datun Road,
Beijing 100012, The People's Republic of China  }
\email{louyq@mail.tsinghua.edu.cn and lou@oddjob.uchicago.edu}
\author{Mei Wu\altaffilmark{3}}
\affil{Particle Astrophysics Laboratory, Institute of High Energy
Physics, Chinese Academy of Sciences, Beijing 100039, The People's
Republic of China}
\author{Jin-Lu Qu\altaffilmark{3}, Shu-Mei Jia\altaffilmark{1}
and Xue-Juan Yang\altaffilmark{1} }

\begin{abstract}
Using the data acquired in the Time To Spill (TTS) mode for long
gamma-ray bursts (GRBs) collected by the Burst and Transient
Source Experiment on board the Compton Gamma Ray Observatory ({\it
BATSE/CGRO}), we have carefully measured spectral lags in time
between the low ($25-55$ keV) and high ($110-320$ keV) energy
bands of individual pulses contained in 64 multi-peak GRBs. We
find that the temporal lead by higher-energy $\gamma-$ray photons
(i.e., positive lags) is the norm in this selected sample set of
long GRBs. While relatively few in number, some pulses of several
long GRBs do show negative lags. This distribution of spectral
lags in long GRBs is in contrast to that in short GRBs. This
apparent difference poses challenges and constraints on the
physical mechanism(s) of producing long and short GRBs. The
relation between the pulse peak count rates and the spectral lags
is also examined. Observationally, there seems to be no clear
evidence for systematic spectral lag-luminosity connection for
pulses within a given long GRB.
\end{abstract}


\section{Introduction}
Temporal delays in the arrival of low-energy photons relative to
that of high-energy photons are well known in the spectra of
gamma-ray bursts (GRBs). Link, Epstein \& Priedhorsky (1993) used
the autocorrelation analysis to investigate temporal properties of
GRBs in different energy bands. Cheng, Ma, Cheng, Lu \& Zhou
(1995) first quantified the time delay of GRBs in the soft energy
band. Fenimore \& Zand (1995) found that the average
autocorrelation of GRB temporal histories is a universal function
that can measure the timescale as a function of energy. The
dependence is a power law in energy with an index of $\sim 0.4$.
This is the first quantitative relationship between temporal and
spectral structures in GRBs. Band (1997) performed a
cross-correlation analysis on a sample of 229 strongest {\it
BATSE} GRBs to demonstrate that the hard-to-soft spectral
evolution is generic in most bursts. Norris, Marani \& Bonnell
(2000) estimated spectral lags between the light curves of 6 GRBs
with known redshifts $z$ in the energy ranges of the {\it BATSE}
channel 3 ($100-300$ keV) and channel 1 ($25-50$ keV), concluding
that the pulse peak luminosity and the spectral lag $\tau_{lag}$
in time anticorrelate with each other and may well be fit with a
power law $L_{53}\approx 1.3(\tau_{lag}/0.01s)^{-1.15}$ where
$L_{53}$ is the GRB luminosity in unit of $10^{53} \hbox{ erg
s}^{-1}$. This appears to be the first valuable although
preliminary information for GRB luminosity based on spectral and
temporal properties of gamma-ray observations alone. Another
tentative GRB luminosity indicator is the empirical relation
between luminosity and variability first proposed by Fenimore \&
Ramirez-Ruiz (2000), indicating a correlation between spectral
lags and \emph{V} (variability) parameter. This correlation was
further demonstrated by Schaefer, Deng \& Band (2001) by
systematically examining the available {\it BATSE} data of
$\tau_{lag}$ and of \emph{V} for 112 GRBs. By extrapolating the
lag-luminosity relation of GRBs (Ioka \& Nakamura 2001), Murakami,
Yonetoku, Izawa \& Ioka (2003) further inferred the star-formation
history in the universe out to redshift $z\sim4$.

The lag-luminosity relation or the variability-luminosity relation
might make GRBs into standard candles as cosmological distance
indicator. This does not mean that all GRBs have the same
luminosity. In this context, let us look at the cases of Cepheid
variables and Type Ia supernovae (SNe Ia). The well-established
period-luminosity relation of Cepheid variables allows us to know
their intrinsic luminosities from periodicities in their light
curves. Similarly, the decline rate of the light curve of a SN Ia
may determine its peak luminosity after some corrections and
adjustments (Phillips 1993; see extensive references in Niemeyer
\& Truran 2000). Now the lag/$V$-luminosity relation might offer a
possibility of estimating a GRB luminosity based on GRB
observations alone. The cosmological significance of this
potential Cepheid-like relation is self-evident. Comparing with
SNe Ia as cosmological distance indicators, GRB cosmology has at
least three apparent advantages (Norris 2003; Dai, Liang \& Xu
2004): (1) much larger redshift $z$ range; (2) much weaker effects
of dust extinction; and (3) weaker possible luminosity evolution
with redshift $z$. Fenimore \& Ramirez-Ruiz (2000) take the lead
in estimating the redshifts $z$ for 220 bright, long duration {\it
BATSE} GRBs by using $V$-luminosity relation. Norris (2002) used a
two-branch lag-luminosity relationship to yield the number
distribution of GRBs in luminosity, distance, and redshift $z$. It
is further inferred that some GRBs are identified to concentrate
near the local galactic superplane, including GRB 980425 which was
known to associate with a supernova. Applying the lag-luminosity
relation to 1218 GRBs with positive lags, hardness ratios, peak
fluxes and durations, Band, Norris \& Bonnell (2003) compiled an
extensive catalog for GRBs redshifts. Dai et al. (2004) attempted
to constrain the mass density of the universe and the nature of
dark energy with a sample of 12 GRBs with known redshifts, peak
energies and break times of afterglow light curves. Their results
are consistent with those from SNe Ia. Undoubtedly, a larger
sample expected from the upcoming {\it SWIFT} satellite may
provide further clues and constraints.

There have been several attempts to interpret the empirical
relations for luminosity versus spectral lag or \emph{V} in GRB
observations. Salmonson (2000, 2001) proposed that these
correlations might be caused by the variety, among GRBs, of
relativistic velocities at which emitting regions move toward the
observer. He introduced the peak number luminosity $N_{pk}$ (i.e.,
photons s$^{-1}$) instead of the peak luminosity $L_{pk}$. With
this characterization, he not only reproduced the result of Norris
et al. (2000), namely $L_{pk}\propto\tau_{lag}^{-1.15}$ but also
obtained a better fit for the equivalent relation
$N_{pk}\propto\tau_{lag}^{-0.98}$. For a burst expanding with a
Lorentz factor $\gamma\gg 1$, he educed
$N_{pk}\propto\tau_{lag}^{-1}$. In terms of the energy
conservation when the radiative cooling dominates, Schaefer (2004)
considered that the average cooling rate per particle in the
emitting region of the jet should be either equal to the total
luminosity $L_{tot}$ over the number of emitting particles
(roughly $\sim M_{jet}/m_{proton}$) or the time derivative of the
mean particle energy $E_{pk}$, that is:
$\dot{E}_{pk}=-L_{tot}/(M_{jet}/m_{proton})$. Using the relation
$L_{tot}=L_{pk}\Omega/(4\pi)$, where $\Omega\approx\pi\gamma^2$ is
the solid angle into which the radiation is beamed at, and
evaluating $\dot{E}_{pk}$ by $[E_{pk}(T_1)-E_{pk}(T_3)]/(T_1-T_3)
\propto\tau_{lag}^{-1}$, they derived the following relation
$L\propto\tau_{lag}^{-1}$. On the other hand, Wu \& Fenimore
(2000) believed that the synchrotron cooling timescale in a
magnetized jet is much shorter than the lag timescale. According
to the internal shock model of GRBs (Rees \& M\'eszaros 1994),
there are three possible sources of time variation structure in
GRB pulses: cooling, hydrodynamics, and geometric angular effects.
Wu \& Fenimore argued that cooling is much too fast to account for
the observed lags and angular effects should be energy
independent. Thus, only hydrodynamical processes are responsible
for these lags. To be more precise, as magnetic fields and
relativistic flows are likely to be involved in various ways,
relativisitic magnetohydrodynamic (RMHD) processes (e.g., Lou
1992, 1993a, 1993b, 1994, 1996, 1998) might be relevant to the
understanding these lags. Ioka \& Nakamura (2001) argued that the
pulse peak luminosity-variability relation might be caused by the
variation in viewing angle of the source jet. The correlation
between pulse lag or luminosity and jet-break time was noticed by
Salmonson \& Galama (2002), first revealing a connection between a
feature of the GRB phase and the afterglow phase. The correlation
may be qualitatively understood from models in which the Lorentz
$\gamma$ factor of an RMHD jet decreases as a function of angle
away from the jet axis. Kocevski \& Liang (2003) established
empirically the connection between the GRB spectral evolution rate
and spectral lag. They suspected that this may eventually reveal
the underlying physical mechanism(s) for the spectral
lag-luminosity correlation.

As has been suspected, the spectral lags of GRBs and their
evolution are vital to probe the physics of GRB. We do want to
know whether the lag-luminosity relation remains to be a universal
feature for GRBs. In fact, Sazonv, Lutovinov \& Sunyaev (2004)
have found that the peak luminosities of GRBs 031203 and 980425
measured from the given cosmological parameter and redshift $z$
are much lower than ones expected with lag-luminosity relation.
Perhaps, only a certain kind of GRBs (e.g., those not associated
with SNe) fit in the lag-luminosity relation. We would also like
to know whether there are cases of negative lags in long GRBs.
Previous observations of spectral lags were usually based on the
data with a relatively low time resolution, such as the data
acquired in the Discriminator Counts (DISCSC) mode of {\it BATSE},
with a time resolution of 64ms. As most spectral lags in time are
fairly small, typically $\leq 100$ms, an analysis of spectral lag
properties using 64ms time bin data would appear somewhat coarse
and insufficient. In contrast, the TTS mode data of {\it BATSE}
provides an opportunity to determine more precise temporal
structures of GRB spectral lags. To resolve more fine spectral lag
information from TTS mode data (e.g. lags less than 64ms) and
especially to examine the distribution and evolution of the
spectral lags are the major tasks we would like to pursue in this
research work.

This paper is structured as follows. Observations and data
analyses are described in \S 2. The results are presented in \S 3.
In \S 4, we summarize the conclusions and provide discussions.

\section{Database }

The {\it BATSE} data acquired at both the TTS mode and the DISCSC
mode were taken in the following four photon energy bands: $25-55$
keV, $55-110$ keV, $110-320$ keV and $>320$ keV. The two data
acquisition modes of operation are fundamentally different: The
TTS mode records the time interval for every 64 $\gamma$-ray
photons, while the DISCSC modes provides the number of
$\gamma$-ray photons in every 64ms time interval.

As we would like to infer and estimate spectral lags with a time
resolution less than 64ms for individual pulses of a GRB, we
interpolate the light curves obtained in the TTS mode with a
temporal subinterval of 8ms resolution under the key assumption
that these 64 photons distribute more or less evenly within each
time interval of TTS mode of operation. The empirical reason of
binning data into 8ms is that the resulting light curve with such
a bin size is fine enough to achieve a reasonable level of
signal-to-noise (S/N) ratio. In order to empirically justify the
validity of such a temporal interpolation scheme, we have first
determined the spectral lags of a dozen GRBs based on the data of
both TTS and DISCSC modes for comparison with time lags longer
than 150 ms. The results are mutually consistent within estimated
error bars.

The procedure of estimating spectral lags of GRBs using a
cross-correlation function (CCF) has been widely adopted (e.g.,
Link et al. 1993; Fenimore et al. 1995; Norris et al. 2000). The
CCF of $x_{1}(t)$ and $x_{2}(t)$ for the time duration of each
pulse of GRBs, where $x_{1}(t)$ and $x_{2}(t)$ are respective
light curves in two different $\gamma-$ray photon energy bands
(namely, energy band I: $25-55$ keV and energy band II: $110-320$
keV), is simply defined by
\begin{equation}\label{1}
\textrm{CCF}(\tau:\nu_1,\nu_2)=\frac{<\nu_1(t+\tau)\nu_2(t)>}
{\sigma_{\nu_1}\sigma_{\nu_2}}\ ,
\end{equation}
where $\nu_i(t)\equiv x_i(t)-<x_i(t)>$ is the light curve of zero
mean and $\sigma_{\nu_{i}}\equiv <\nu_{i}^{2}>^{1/2}$ is the
standard deviation away from the mean. The spectral lag time
$\tau$ is defined as such to maximize the CCF($\tau;\nu_1,\nu_2$).
A positive $\tau$ corresponds to an earlier arrival or leading
higher-energy $\gamma-$ray photons. In order to reduce scattering
of noises in the CCF, we use a Gaussian function with an
additional linear term to fit the part around the peak of the CCF;
in comparison, Norris et al. (2000) used quadratic as well as
cubic fits. The fitting range is chosen based on (1) a
sufficiently long length such that fits are always convergent and
(2) a reasonably short length such that peaks are well identified
and fit by relatively simple functions. Uncertainties of lags are
determined by estimating errors transferred via the fitting
parameters. As the fitting code can provide a $1\sigma$ parameter
uncertainties, we calculate $\tau\prime$ for the fitting curves
with parameters adding or subtracting $1\sigma$ uncertainties. The
error in $\tau$ is thus defined as the maximum of the absolute
difference between $\tau$ and $\tau\prime$. On the theoretical
ground, error bars should be greater than the time resolution of
the data, that is, greater than 8ms and 64ms for the TTS and
DISCSC data, respectively. Spectral lags calculated for the TTS
and DISCSC data of a dozen GRBs with lags $>150$ms do show
consistency within estimated error bars.

\begin{figure}[htbp]
  \includegraphics[bb=0 0 320 245, width=12cm]{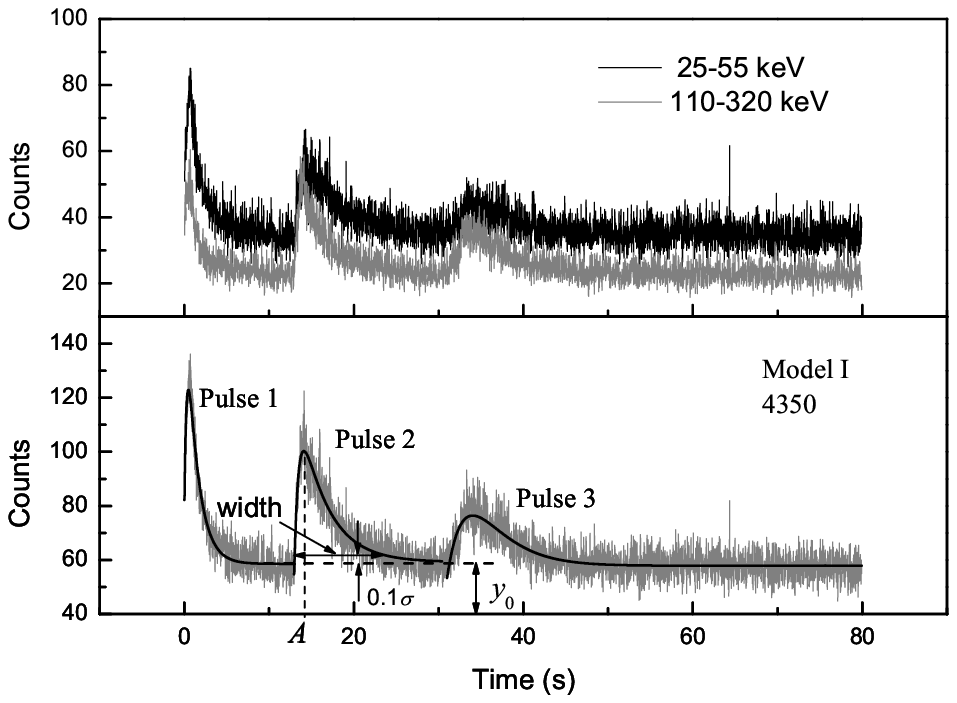}
  \caption{Upper panel: Two light curves of
$\gamma-$ray photon energy bands I and III of GRB 4350. Three
pulses can be clearly identified in this GRB event and the higher
energy $\gamma-$ray photons lead the lower ones (i.e., positive
spectral lags). Lower panel: Each pulse may be well fit by Model
I. The procedure of determining the width of a pulse is also shown
here. }
  \label{fig1}
\end{figure}

The determination of a pulse duration time is equivalent to an
estimation of the pulse width. However, it is not trivial to
approach this problem simply by a code simulation. While the pulse
temporal evolution of a GRB has often been described by a fast
rise followed by an exponential decay (FRED) profile (e.g.,
Fishman et al. 1994; Fenimore 1999 ), there are other fitting
profiles such as stretched exponentials, Gaussian (Norris et al.
1996), power-law decays (Ryde \& Svensson 2000) or even some
irregular shapes. For those GRB pulses with relatively `standard'
profile, we adopt the following model to fit their light curves,
namely
$$
\textrm{Model I}:\qquad\quad y(t)=y_0+A\bigg\{
\exp\bigg[\frac{-(t-t_{max})}{\sigma_r}\bigg]
-\exp\bigg[\frac{-(t-t_{max})}{\sigma_d}\bigg]\bigg\}\ ,
$$
where $y_0$ is a background offset, $t_{max}$ is the time of the
pulse's maximum intensity, $A$; $\sigma_r$ and $\sigma_d$
represents the exponential rise ($t<t_{max}$) and decay
($t>t_{max}$) time constants, respectively [see eq. (1) in Norris
et al. 1996], and
$$
\textrm{Model II:}\qquad\qquad y(t)=\frac{A}{w(\pi/2)^{1/2}}
\exp\bigg[\frac{-(t-t_{max})^{2}}{2w^{2}}\bigg]+a+bt+ct^{2}\ , \ \
\qquad\qquad
$$
where $y(t)$ is a Gaussian function with a quadratic term
representing a background, $t_{max}$ and $A$ are similar to those
in Model I and $w$, $a$, $b$, $c$ are four fitting parameters.
Model I has been applied for a GRB with a FRED profile, while
Model II is more suitable for a GRB with a symmetric profile. In
these two types of profile models, a GRB profile can be seen as a
peak component superposed onto a background of either a constant
$y_0$ (Model I) or a quadratic fit $a+bt+ct^{2}$ (Model II). With
a statistical weight of $w_{i}=1/y_{i}$, we fit the TTS data for
GRBs, taking the pulse width measured $0.1\sigma$ above the
background ($y_0$ or $a+bt+ct^{2}$) (see Fig.\ref{fig1}). For
those GRB pulses of irregular shapes, we determine their pulse
widths by direct visual examinations.

\section{Results of Data Analysis}

\subsection{Distribution of Spectral Lags in Time}

Several thousands of GRBs have been observed by {\it BATSE/CGRO}.
As we plan to analyze the evolution of pulse spectral lags of long
GRBs, we chose those GRBs with well separated multi-pulse profiles
and good S/N ratios. A sample of 64 multi-peak long GRBs have been
carefully selected from the TTS data. We calculate the spectral
lags in time for every individual pulse of the selected sample.
From the left to right columns in order, Table 1 lists the GRB
trigger number according to the `{\it BATSE} Catalog', the numeral
identification of individual pulse in each GRB event, the pulse
peak time (or position) of the low-energy band $25-55$ keV, the
peak count rate of the low-energy band $25-55$ keV, the pulse peak
time (or position) of the high-energy band $110-320$ keV, the peak
count rate of the high-energy band $110-320$ keV, and the inferred
spectral lags in time, where the peak count rate is determined by
the average of three bins around the maximum. Due to the presence
of a Poisson background variation in the light curve, we note that
the sign of a lag does not necessarily always accord with that of
the difference between the peak times of high and low energy
bands.

The histogram of spectral lags for 341 pulses in 64 GRBs has a
distinctly asymmetric sharp peak-like distribution as displayed in
Figure 2. Also seen from Figure 2 is the important fact that
shorter lags are much more numerous than longer lags. A large
majority of spectral lags clearly show earlier arrivals of
$\gamma-$ray photons in the high-energy band, with the maximum lag
distribution at $\tau\sim 30$ms. In contrast, only some 20 long
GRBs show negative lags for which the absolute values are greater
than their errors. Among these cases, there are 10 and 5 lags
reaching $2\sigma$ and $3\sigma$ significant levels, respectively.
While most spectral lags are positive, those negative spectral
lags, if further confirmed to be real, would become additional
constraints on physical models for GRBs. We have carefully
examined those negative lags greater than $2\sigma$ significant
level one by one to make sure that such negative lags are not
caused by pulse overlaps. There are two major situations where
negative spectral lags may arise. One situation usually occurs at
the beginning of a GRB, such as the two GRBs 6333 and 7247, while
the other situation happens as spectral lags vary gradually from
positive to negative ones, such as the GRBs 5773 (see Fig. 3),
7277 and 7301. The GRB 6672 is an exception: it began with a
positive spectral lag, then negative lags appeared, but returned
to positive lags again in the end. Sometimes, the lag of the first
pulse in a GRB cannot be well determined for the trigger time is
not really at the beginning of the burst.

Gupta, Gupta \& Bhat (2002) have studied spectral lags for GRBs of
short durations. They observed in a sample of 156 {\it BATSE} GRBs
with T90 (i.e., $90\%$ of the duration time of a GRB) less than
2s. Unlike GRBs of longer durations in our data analysis, they
found the percentage of short GRBs with negative $\tau_{lag}$ is
$\sim 26\%$, which is much higher than our result -- only 10 and 5
negative spectral lags reaching above $2\sigma$ and $3\sigma$
level respectively among 341 pulses. Meanwhile, over 70\% of short
GRBs in their data sample have positive spectral lags in pulses.
These significantly different distributions in positive and
negative pulse spectral lags imply that there might be entirely
different physical mechanisms responsible for long and short GRBs.
It should be clearly noted that in estimating spectral time lags,
there is an algebraic sign difference between our notation and
that of Gupta et al. (2002) (i.e., our positive lags are their
negative lags and vice versa).

\subsection{Temporal Evolution of Spectral Lags}

We can hardly see any systematic evolution of spectral lags. In
other words, GRB spectral lags differ from each other in great
varieties. The spectral lags of most GRBs do not change in the
order of magnitude. However, a few of them varied significantly
(e.g., GRBs 5773 and 6672). Some GRBs show their spectral lags
increase (e.g., GRBs 4350 and 5548) or decrease (e.g., GRB 5773)
regularly with increasing time. Considering the different physical
mechanisms for GRBs (e.g., Piran 1999; M\'eszaros 2002), this
dissimilarity in the evolution of GRB lags is not strange.

To give a graphic description, we present in Figure 4 spectral
lags versus peak count rates for those long GRBs with more than 9
pulse peaks, where the count rates are the sum of those in the
energy bands I and III.

As shown by Figure 4, relations between the count rates and the
spectral lags are diverse. For example, in the case of GRB 7113,
the spectral lags remain almost constant for different count
rates. On the other hand, the spectral lags in GRBs 6124, 6587,
7277 and 7975 increase with decreasing count rates. Counter
examples of such trend of variation do exist: the spectral lags of
GRBs 7605 and 7954 tend to increase with increasing count rates.
There seems to be no apparent relation between spectral lags and
luminosity in general for pulses within a given long GRB.

\section{Conclusions}

We have calculated the spectral lags in time in pulses of 64 long
GRBs using the TTS data of the {\it BATSE/CGRO} in the two
$\gamma-$ray photon energy bands $25-55$ keV and $110-320$ keV.
For the majority of pulses in GRBs, we see clear signals of
earlier arrivals of high-energy $\gamma-$ray photons, with a
spectral lag distribution peaked at a lag time of $\tau\sim 30$ms.
However, a few pulses do show negative lags, that is, lower-energy
$\gamma-$ray photons arrived earlier than higher-energy
$\gamma-$ray photons for pulses in a GRB event. While we cannot
provide a rigorous statistical test for these negative lags as we
do not know the actual distribution of spectral lags, the
existence of such negative spectral lags in time for pulses in
GRBs appears to be significant enough.

The dependence of the peak count rates on spectral lags varies
significantly on the basis of individual GRB event.

Several GRBs show that their spectral lags either increase or
decrease regularly with increasing time. Wu \& Fenimore (2000)
argued that time lags between different energy $\gamma-$ray
photons might be mainly determined by the relevant dynamical
timescale. Such regular and systematic changes in a GRB may be
caused by some special processes associated with the shell ejecta
movement. Based on our data analysis, we note that there is no
strong evidence for a general correlation between spectral lags
and luminosities to hold for pulses within a given GRB event.
Although only one case, Hakkila \& Gilblin (2004) noted that the
peak luminosity ratio between two peaks of GRB 5478 is not in
agreement with the ratio predicted by the lag versus peak
luminosity.

\acknowledgments We are grateful to the anonymous referee for a
careful reading of several versions of this manuscript and for
encouragement, constructive criticisms, valuable suggestions and
helpful comments to improve the manuscript. This work has been
partially supported by the National Science Foundation of China
(NSFC 10273010) and by the State Basic Science Research Projects
of China (TG20000776). Y.Q.L. has been supported in part by the
ASCI Center for Astrophysical Thermonuclear Flashes at the Univ.
of Chicago under Department of Energy contract B341495, by the
Special Funds for Major State Basic Science Research Projects of
China, by the THCA, by the Collaborative Research Fund from the
National Natural Science Foundation of China (NSFC) for Young
Outstanding Overseas Chinese Scholars (NSFC 10028306) at the NAOC,
Chinese Academy of Sciences, by NSFC grant 10373009 at the
Tsinghua Univ., and by the Yangtze Endowment from the Ministry of
Education at the Tsinghua Univ. Affiliated institutions of Y.Q.L.
share this contribution.

\begin{figure}[htbp]
  \includegraphics[bb=0 0 320 245, width=12cm]{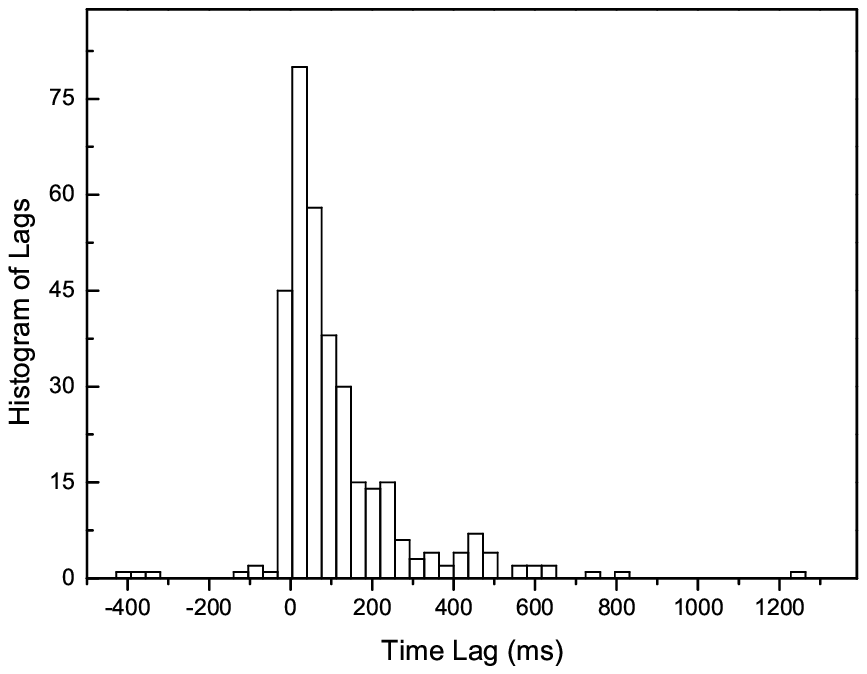}
  \caption{The histogram for spectral lags of all
341 pulses in 64 GRBs of {\it BATSE/CGRO} data acquired in the TTS
mode. Positive spectral lags indicate earlier arrivals of
higher-energy $\gamma$-ray photons. The two energy bands of
$\gamma$-ray photons are $25-55$ keV and $110-320$ keV,
respectively. }
  \label{fig2}
\end{figure}

\begin{figure}[htbp]
  \includegraphics[bb=0 0 320 245, width=12cm]{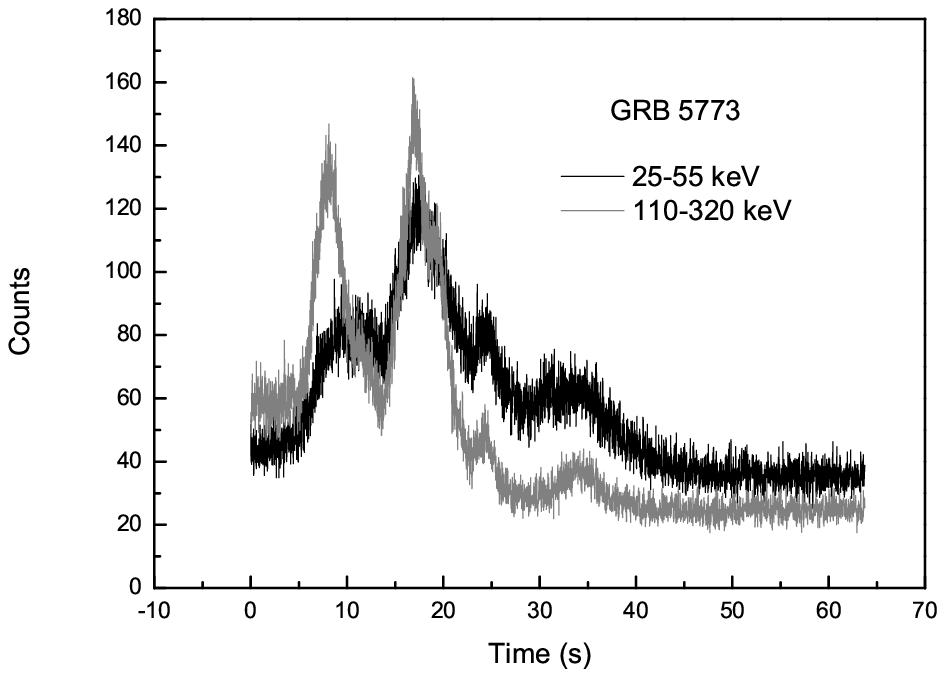}
  \caption{Two light curves of $\gamma-$ray photons
in energy bands I and III of GRB 5773. Four pulses can be readily
identified in this GRB event. The spectral lags change from
positive to negative as can be seen by a direct visual inspection.
}
  \label{fig3}
\end{figure}

\begin{figure}[htbp]
  \includegraphics[bb=0 0 320 245, width=12cm]{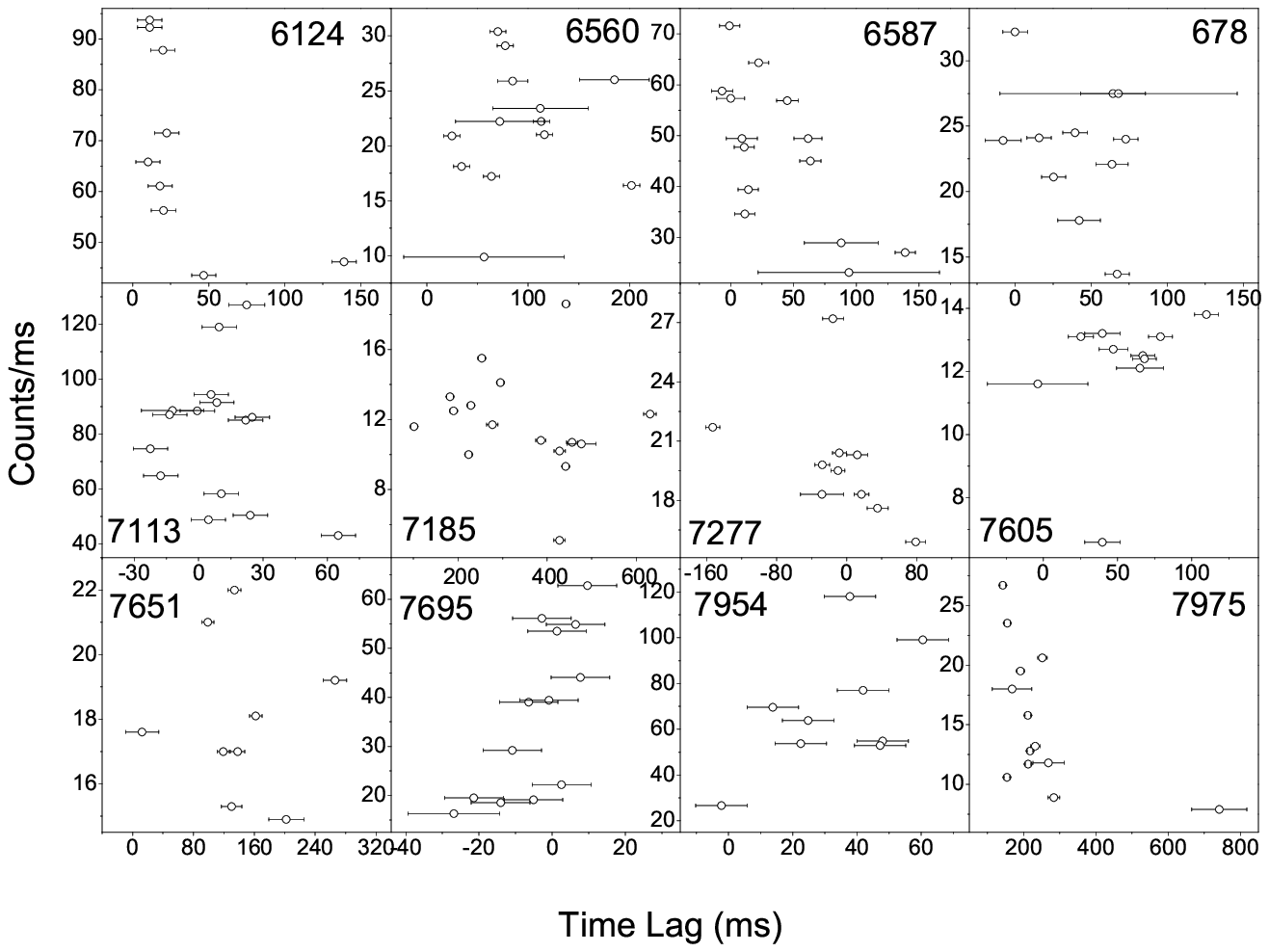}
  \caption{Variations of count rates versus spectral
lags in time for 12 multi-pulse GRBs of {\it BATSE/CGRO} in the
TTS mode. The numeral marked at one appropriate corner in each
panel is the trigger number of that GRB given by `The {\it BATSE}
Catalog Burst Name'. }
  \label{fig4}
\end{figure}

\begin{deluxetable}{crrrrrrr}
\tabletypesize{\scriptsize} \tablecaption{Spectral Lags of GRBs of
{\it BATSE/CGRO} in the TTS Mode \label{tbl-2}} \tablewidth{0pt}
\tablehead{ \colhead{GRB Trigger}& \colhead{Pulse }&\colhead{Peak
1}&\colhead{Peak Rate 1}&
\colhead{Peak 2}&\colhead{Peak Rate 2}&\colhead{Spectral Lag}\\
\colhead{Number}&\colhead{ID (i)}&\colhead{Position (s)}&
\colhead{(counts/s)}&\colhead{Position (s)}&
\colhead{(counts/s)}&\colhead{in Time (ms)}}\startdata
143 &   1   &   1.26&   46704.09  &  1.22&  137143.75&   60.8$\pm$   11.2  \\
    &   2   &   2.39&   43109.49  &  2.32&   116417.9&   44.8 $\pm$   8.96 \\
    &   3   &   4.62&   35436.35  &  3.72&   92801.71&   49.6 $\pm$   8   \\
    &   4   &   5.13&   29438.69  &  5.1 &   77592.85&   32.64$\pm$   8   \\
    &   5   &   7.7 &   18834.17  &  7.7 &   27545.6 &   27.2 $\pm$   8   \\
    &   6   &   8.78  &   23030.62& 8.77 &  37565.68 &     32 $\pm$   8   \\
    &   7   &   48.46 &   20298.77& 48.42&  33988.35 &   51.2$\pm$    8   \\
    &   8   &   53.09 &   12500.69& 52.74&  13794.32 &   112.64$\pm$ 8.96  \\  \hline
219 &   1   & 113.71  &   9427.04 &113.98&  13191.89 &   448 $\pm$   54.4  \\
    &   2   & 116.17  &   20678.17&115.72&   41205.2 &   471.36 $\pm$   54.08   \\
    &   3   & 119.91  &   14846.08&119.35&  21079.44 &   478.4$\pm$ 15.6  \\
    &   4   & 127.76  &   8229.48 &127.52&   6275.81 &   438.08$\pm$ 57.2 \\
    &   5   & 131.93  &   7239.85 & 132.04  &5768.08 &   424 $\pm$   16  \\
    &   6   & 134.15  &   8348.71 & 133.82  &   6385.35 & 435.2$\pm$ 12.96 \\  \hline
222 &   1   &   0.25  &   5444.67 & 0.54    &   5325.27 & 82.66$\pm$ 22.04 \\
    &   2   &   1.42  &   6164.05 & 1.33    &   6051.12 & 127.44$\pm$104.04  \\
    &   3   & 61.98   &   5371.59 & 62.23   &   4506.68 & 263.2 $\pm$   28  \\
    &   4   & 69.05   &   7433.04 & 68.75   &   9036.54 & 234.75$\pm$   8   \\  \hline
249 &   1   &   15.3  & 14463.06  &   14.23 &   18077.19& 64.64 $\pm$ 10.24 \\
    &   2   &   18.98 & 27320.19  &   19.02 &   60113.95&   -5.76 $\pm$   8   \\
    &   3   &   21.96 & 39382.03  &   21.26 &   93848.28&   107.52$\pm$   8   \\
    &   4   &   25.02 & 26597.9   &   25.02 &   43470.93&   33.28 $\pm$   16.64   \\
    &   5   &   32.24 & 13462.76  &   32.1  &   13356.52&   120 $\pm$  12  \\
    &   6   &   41.48 & 9394.71   &   42.68 &   8170.37 &   252.8$\pm$   8.8 \\  \hline
612 &   1   &   5.16  & 4827.79   &   5.34  &   5830.81 &   144 $\pm$ 16.8    \\
    &   2   &   8.35  & 5837.57   &   7.84  &   7831.93 &   102.4 $\pm$ 19.2    \\
    &   3   &   10.71 & 5723.82   &   10.42 &   6294.77 &   308.42  $\pm$   32.86   \\  \hline
678 &   1   &   0.94  &   7418.09 &   0.87  &   20105.52&  64.29$\pm$   21.25   \\
    &   2   &   0.94  &   7418.09 &   0.87  &   20105.52&  68  $\pm$   78  \\
    &   3   &   1.51  &   8440.53 &   1.42  &   23727.88&  0$\pm$   8.13    \\
    &   4   &   4.72  &   7413.53 &   4.7   & 16477.24  &  -7.87$\pm$ 11.81   \\
    &   5   &   6.04  &   7125.27 &   5.99  &   13956.33&   25.4 $\pm$ 8   \\
    &   6   &   6.65  &   7897.96 &   6.53  &   16235.25& 15.75 $\pm$  8   \\
    &   7   &   7.55  &   7274.85 &   7.7   &   14807.79& 63.74 $\pm$ 10.62   \\
    &   8   &   9.39  &   8316.6  &   9.34  &   16141.58& 39.51 $\pm$   8   \\
    &   9   &   10.91 &   7915.53 &   10.82 &   16109.92& 72.8  $\pm$   8   \\
    &   10  &   13.38 &   6987.72 &   13.47 &   10829.17& 42.05 $\pm$ 14.02 \\
    &   11  &   18.27 &   6483.86 &   18.18 &   7221.36 & 67.2  $\pm$  8  \\  \hline
841 &   1   &   2.18  &   6347.55 &   1.89  &   3711.56 & 80.64 $\pm$   8   \\
    &   2   &   12.28 &   8788.98 &   12.19 &   9610.47 & 76.74 $\pm$   8   \\
    &   3   &   15.01 &   6796.78 &   13.98 &   5130.81 & 117.6 $\pm$   8.4 \\
    &   4   &   16.99 &   5371.46 &   16.78 &   3672.38 &151.2  $\pm$   11.2    \\  \hline
869 &   1   &   16.95 &   4094.31 &   17.98 &   5931.28 &142.53 $\pm$   16.77   \\
    &   2   &   96.1  &   5089.73 &   94.95 &   6021.75 &117.6  $\pm$   19.6    \\
    &   3   & 110.78  &   5478.87 &   110.97&   7979.88 &   336 $\pm$   16.8    \\  \hline
973 &   1   &   3.02  &   8807.73 &   2.18  &   12537.31& 651.2 $\pm$   10.4    \\
    &   2   &   23.41 &   7051.8  &   24.56 &   7036.9  &   112 $\pm$   11.2    \\  \hline
1025&   1   &   0.06  &   6741.27 &   0.24  &   6313.29 & 60.16 $\pm$   33.12   \\
    &   2   &   1.65  &   10658.4 &   1.65  &   13611.31& 5.28  $\pm$   8   \\
    &   3   &   2.13  &   14688.1 &   2.02  &   23291.17& 70.4  $\pm$   12.8    \\ \hline
1085&   1   &   3.9   &   13195.1 &   1.83  &   29906.9 &474.24 $\pm$   74.24   \\
    &   2   &   6.5   &   24122.74&   5.75  &   41496.01&   752 $\pm$   46  \\
    &   3   &   7.6   &   19972.91&   7.39  &   20160.15& 172.8 $\pm$   44.8    \\ \hline
1190&   1   &   0.17  &   4890.6  &   0.11  &   5134.78 & 28.8  $\pm$   26.4    \\
    &   2   &   0.34  &   5561.23 &   0.31  &   5340.42 & 27.58 $\pm$   28.42   \\
    &   3   &   0.82  &   5444.33 &   0.87  &   6402.98 & 28.0  $\pm$   8   \\
    &   4   &   2     &   5080.79 &   1.9   &   5751.9  & 33.6  $\pm$   8   \\
    &   5   &   2.92  &   4213.65 &   2.86  &   3724.16 & 29.5  $\pm$   8   \\  \hline
1204&   1   &  0.09   &   5554.46 &   0.12  &   9852.05 & -0.96 $\pm$   8   \\
    &   2   &  0.99   &   4407.6  &   1     &   4956.15 & 7.87  $\pm$   8   \\
    &   3   &  2.98   &   4298.2  &   2.9   &   3788.45 & 78.91 $\pm$ 11.9  \\  \hline
4350&   1   &  0.64   &   9804.3  &   0.71  &   7391.88 &   196 $\pm$ 14  \\
    &   2   & 17.13   &   7058.09 & 13.69   &   7101.72 & 344.8 $\pm$ 17.28 \\
    &   3   & 34.59   &   6380.98 & 33.67   &   4934.64 &   416 $\pm$ 56  \\  \hline
5447&   1   & 6.59    &  10676.96 &   6.9   &   5087.84 &   96  $\pm$ 9.6 \\
    &   2   & 13.63   &   7045.25 & 13.25   &   3737.58 & 207.2 $\pm$ 14.4 \\  \hline
5548&   1   & 16.87   &   8533.64 & 16.9    &   10107.59&  16.0 $\pm$  8 \\
    &   2   & 51.14   &   6082.37 & 50.42   &   6166.6  &159.04 $\pm$ 10.08 \\
    &   3   & 101.5   &   4548.6  & 101.51  &   3138.83 & 254.4 $\pm$ 25.2   \\
    &   4   & 163.47  &   4564.25 & 166.1   &   3556.31 & 358.4 $\pm$ 12.8   \\  \hline
5568&   1   & 0.75    &   5865.03 & 0.86    &   8639.73 & 82.66 $\pm$ 39.95   \\
    &   2   &   1.9   &   11073.6 & 1.83    & 34474.03  & 62.88 $\pm$ 48.96   \\
    &   3   & 2.75    &  12678.03 & 2.74    & 32657.64  & -10.4 $\pm$ 8  \\ \hline
5572&   1   & 0.17    &   11380.6 & 0.02    & 14793.24  & -1.92 $\pm$  8   \\
    &   2   &   0.77  &   15103.77& 0.76    & 28440.03  & 5.64  $\pm$  8   \\
    &   3   & 4.22    &   11405.69& 4.54    & 10612.42  & -3.36 $\pm$ 15.12   \\  \hline
5575&   1   &   0.3   & 11065.87  & 0.23    & 15611.22  &144.64 $\pm$ 13.44   \\
    &   2   &   6.68  &   6076.13 & 6.02    &   5046.16 &   96  $\pm$ 16  \\
    &   3   &   8.68  &   10190.02& 8.58    &   11250.84& 44.35 $\pm$  8   \\
    &   4   &   9.22  &   11655.85& 9.25    &   9916.92 & 72.11 $\pm$  8   \\
    &   5   &   9.96  &   11351.2 & 9.93    &   10029.7 &   80  $\pm$ 20.8    \\  \hline
5591&   1   &   51.22 &   5122.53 & 50.76   &   5015.69 & 289.6 $\pm$ 12  \\
    &   2   & 100.25  &   5690.69 & 99.92   &   5648.2  & 185.6 $\pm$ 9.6 \\
    &   3   & 144.13  &   6561    & 143.83  &   5631.88 &   240 $\pm$ 24  \\  \hline
5601&   1   & 4.47    &   5476.04 & 2.37    &   5029.91 & 563.6 $\pm$ 96.6 \\
    &   2   &   7.4   &   7002.75 & 7.13    &   9498.09 & 460.1 $\pm$ 55.8 \\  \hline
5621&   1   &   3.49  & 11212.87  &   3.49  & 14532.77  & 29.44 $\pm$   8   \\
    &   2   &   3.92  &   19020.7 &   3.91  & 36243.95  & 54.5  $\pm$   8   \\
    &   3   &   5.74  & 12989.42  &   5.74  &   17888.9 & 25.6  $\pm$   8   \\
    &   4   &   6.91  & 12021.35  &   6.85  & 13601.25  & 12.7  $\pm$   8   \\
    &   5   &   7.14  & 13461.48  &   7.18  & 22535.45  & 60.48 $\pm$   8   \\
    &   6   &   9.98  &   9988.35 &   9.95  &   9078.3  & -2.4  $\pm$   8   \\  \hline
5628&   1   &   6.91  &   7630.21 &   7.18  &   9852.22 & -6.4  $\pm$   43.2    \\
    &   2   &   7.7   &   8090.29 &   7.71  & 11792.75  &   40  $\pm$   8   \\
    &   3   &   9.03  &   8091.89 &   9.01  &  10009    & -2.82 $\pm$   8   \\
    &   4   &   10.06 &   10466.05&   10.2  &  15583.61 & 21.7  $\pm$   8   \\  \hline
5654&   1   &   21.18 &   6542.94 &   14.62 &   9984.46 & 499.2 $\pm$ 76.8  \\
    &   2   &   33.65 &   5905.52 &   37.66 &   6685.05 &167.84 $\pm$ 23.92 \\
    &   3   &   78.34 &   4844.36 &   77.74 &   3532.31 &104.8  $\pm$ 61.2  \\
    &   4   &   92.96 &   4401.31 &   92.72 &   3812.54 & 345.6 $\pm$ 25.6  \\  \hline
5726&   1   &   3.47  &   6173.52 &   3.38  &   4185.61 &   112 $\pm$ 14   \\
    &   2   &   5.38  &   6131.64 &   5.03  &   6228.25 & 62.09 $\pm$  8   \\
    &   3   &   7.37  &   7520.29 &   6.98  &   7547.79 & 36.48 $\pm$  8   \\  \hline
5729&   1   &   28.41 &   7541.93 &   27.98 &   4970.05 &372.74 $\pm$ 13.31 \\
    &   2   &   38.68 &   8518.49 &   36.78 &   5705.94 & 585.73$\pm$ 39.94 \\  \hline
5731&   1   &   14.25 &   6106.97 &   14.03 &   6032.85 & 89.86 $\pm$  8   \\
    &   2   &   34.31 &   6040.33 &   34.12 &   7833.2  &   84  $\pm$   8   \\
    &   3   &   38.69 &   6479.97 &   38.58 &   7496.68 & 99.2  $\pm$   8   \\  \hline
5773&   1   &   11.34 & 11345.14  &   8.11  &  18127.99 & 1256.2$\pm$ 78.4 \\
    &   2   &   17.72 & 16424.96  & 16.87   & 19671.89  & 195.2 $\pm$ 15.6 \\
    &   3   &   23.52 & 11741.22  & 24.32   &   7372.24 & 140.2 $\pm$ 25.4 \\
    &   4   &   33.66 &   8780.13 & 35.18   &   5233.09 &-364.42$\pm$ 14.02 \\  \hline
5989&   1   &   0.38  & 69045.97  &  0.27   & 110138.38 & 70.4  $\pm$ 8   \\
    &   2   &   1.1   & 21951.07  & 1.03    &   8125.8  & 68.88 $\pm$ 8   \\
    &   3   &   18.7  & 15184.64  & 18.58   &   7545.06 & 39.51 $\pm$ 8   \\
    &   4   &   20.48 & 28495.33  & 20.36   & 11093.89  & 64.92 $\pm$ 8   \\
    &   5   &   23.78 & 30222.18  & 23.54   &   7458.01 & 78.72 $\pm$ 8   \\  \hline
6100&   1   &   6.37  & 11256.17  & 6.21    & 21148.39  & 84.1  $\pm$ 10.51 \\
    &   2   &   8.31  &15987.26   & 8.31    & 33571.85  & 69.76 $\pm$   8   \\  \hline
6124&   1   &   1.18  & 14047.67  & 1.14    & 32134.29  & 139.2 $\pm$   8   \\
    &   2   &   6.12  & 18155.54  &   6.1   & 42918.34  & 18.05 $\pm$   8   \\
    &   3   &   7.34  & 17233.61  &  7.08   &   39099   & 20.4  $\pm$   8    \\
    &   4   &   8.22  & 16553.99  &  8.64   & 54925.32  & 22.5  $\pm$   8   \\
    &   5   &   9.34  & 21629.41  &  9.42   & 66180.71  & 19.76 $\pm$   8   \\
    &   6   &   10.02 & 19731.38  & 10.02   & 72543.17  & 11.25 $\pm$   8   \\
    &   7   &   11.3  & 16476.98  & 11.29   & 49369.37  & 10.08 $\pm$   8   \\
    &   8   &   12.77 & 21607.11  & 12.76   & 72135.56  & 11.2  $\pm$   8   \\
    &   9   &   14.86 & 15829.08  & 14.78   & 27685.85  & 46.77 $\pm$   8   \\  \hline
6157&   1   &   0.94  &   6198.38 &  0.89   &   8506.09 & 35.58 $\pm$   8   \\
    &   2   &   2.27  &   6522.94 &  2.13   & 11378.44  & 45.63 $\pm$ 14.88 \\
    &   3   &   2.54  &   6661.34 &  2.59   & 10456.45  & 46.77 $\pm$  9.74 \\
    &   4   &   3.39  &   6029.75 &  3.42   & 11717.95  & 27.78 $\pm$   8   \\
    &   5   &   4.46  & 10397.08  &  4.48   &  20453.35 & 18.08 $\pm$   8   \\
    &   6   &   5.91  &   6587.55 &  6.06   & 13462.95  & 39.36 $\pm$   8   \\
    &   7   &   6.39  &   5681.97 &   6.4   & 10786.04  & 0.57  $\pm$142.91 \\
    &   8   &   7.21  &   9376.81 & 7.18    &   19577   & 63.17 $\pm$   8   \\
    &   9   &   8.8   &   5467.62 &   8.81  &   5430    & 44.44 $\pm$   8   \\
    &   10  &   9.15  &   6066.33 &   9.11  & 11111.71  & 16.93 $\pm$   8   \\  \hline
6168&   1   &   25.2  &   9263.25 &   25.11 & 15846.34  &   144 $\pm$ 16.8    \\
    &   2   &   27.54 & 12669.29  &   27.27 & 30621.96  & 108.99$\pm$ 41.92   \\  \hline
6333&   1   &   1.54  &   6801.92 &   1.89  &   4335.05 &-347.12$\pm$ 26.52   \\
    &   2   &   4.97  &   7257.69 &   5.01  &   5313.64 & 43.39 $\pm$  8   \\
    &   3   &   6.39  &   7056.58 &   6.43  &   4654.19 & 3.36  $\pm$ 11.76 \\  \hline
6336&   1   &   4.36  &   9241.58 &   4.46  &   19402.64&  18.24$\pm$ 21.89 \\
    &   2   &   5.4   & 13560.22  &   5.18  & 29018.18  & 239.68$\pm$ 15.84 \\
    &   3   &   6.71  & 11840.69  &   6.38  & 12467.35  & -11.29$\pm$ 14.11 \\  \hline
6404&   1   &   3.48  & 19305.13  &   3.43  &   36109.4 & 246.4 $\pm$   8   \\
    &   2   &   8.73  & 25334.39  &   8.54  &   34959.45& 123.2 $\pm$ 12  \\
    &   3   &   38.42 &   6196.4  & 38.06   &   4875.59 & 229.6 $\pm$ 19.6 \\
    &   4   &   40.43 &   6802.02 & 40.15   &   5109.6  & 209.98$\pm$ 18.53 \\  \hline
6554&   1   &   0.62  &   5432.32 &   0.23  &   4483.81 & 601.6 $\pm$ 33.6  \\
    &   2   &   8.9   &   6558.73 &   9.02  &   3394.91 & 14.02 $\pm$   8   \\  \hline
6560&   1   &   0.56  &   8513.42 &   0.17  &   9543.97 & 34.4  $\pm$   8   \\
    &   2   &   1.21  &   6045.44 &   1.48  &   3818.73 & 56.45 $\pm$ 79.03 \\
    &   3   &   4.19  &   7594.62 &   4.14  &   9639.94 & 63.84 $\pm$  8   \\
    &   4   &   16.98 &   8918.76 &   17.28 &  13295.21 & 112.9 $\pm$  8   \\
    &   5   &   17.58 &   8604.36 &   17.34 & 17355.12  &185.15 $\pm$  34.27   \\
    &   6   &   17.87 &   8850.72 &   17.77 & 14587.61  & 112.32$\pm$ 47.04   \\
    &   7   &   22.7  &   9828.27 &   22.95 & 12401.11  &   72  $\pm$ 43.68   \\
    &   8   &   23.5  & 11299.79  &   23.35 & 19121.28  & 70.16 $\pm$   8   \\
    &   9   &   24.25 & 10132.65  &   24.46 & 15807.05  & 84.67 $\pm$ 14.78   \\
    &   10  &   25.89 &12569.71   &   25.84 & 16564.29  & 77.38 $\pm$   8   \\
    &   11  &   29.08 & 11165.58  &   29.02 &   9849.95 &115.72 $\pm$   8   \\
    &   12  &   35.13 &   7388.69 &   35    &   9025.26 & 202.24$\pm$   8   \\  \hline
6587&   1   &   10.13 & 9493.22   &   9.7   & 19390.77  &   88  $\pm$ 29.44   \\
    &   2   &   11.25 & 11099.49  &   10.92 & 15888.32  & 138.69$\pm$  8   \\
    &   3   &   11.87 & 11200.93  &   11.7  & 11877.43  & 94.08 $\pm$ 72.24   \\
    &   4   &   13.81 & 14157.62  &   13.58 & 30840.62  & 63.36 $\pm$  8.45    \\
    &   5   &   14.97 &   12111.7 &   15.05 &   27281.2 & 14.11 $\pm$  8   \\
    &   6   &   16.81 &  14243.11 &  16.94  & 35134.52  & 61.6  $\pm$ 11.2    \\
    &   7   &   21.5  & 15612.32  & 20.68   &   41642.4 &   0   $\pm$   11.23   \\
    &   8   &   22.69 & 11547.73  & 22.73   & 36169.06  &  10.9 $\pm$   8   \\
    &   9   &   23.21 & 16687.85  & 23.18   &   42142.4 & -6.72 $\pm$   8.4 \\
    &   10  &   24.78 & 17749.89  & 24.82   &  46530.71 & 22.4  $\pm$   8   \\
    &   11  &   26.11 & 13593.47  & 26.35   & 35845.33  &   9   $\pm$   12.37   \\
    &   12  &   27.53 & 18791.42  & 26.98   & 52804.52  & -0.8  $\pm$   8   \\
    &   13  &   29.99 & 15973.07  & 29.94   & 40879.16  & 45.16 $\pm$   8.47    \\
    &   14  &   31.15 & 12667.77  & 31.1    &   21965.7 & 11.29 $\pm$   8   \\  \hline
6672&   1   &   0.93  & 10997.34  & 0.87    &   9432.11 & 120.22$\pm$   9.25    \\
    &   2   &   3.88  &   8699.88 & 3.86    &   4635.64 & -99.2 $\pm$   35.2    \\
    &   3   &   7.25  &   8011.5  & 6.85    &   4869.85 &  27.84$\pm$   9.28    \\  \hline
7113&   1   &   3.37  & 14402.16  & 3.28    & 28654.04  & 65.09 $\pm$   8   \\
    &   2   &   6.51  & 17721.43  & 7.42    & 32685.95  & 24.1  $\pm$   8   \\
    &   3   &   8.78  & 17510.57  & 8.74    & 31280.82  & 4.51  $\pm$   8   \\
    &   4   &   12.03 & 19428.95  & 10.26   & 38888.76  & 10.54 $\pm$   8    \\
    &   5   &   16.69 & 23126.18  & 16.71   & 65265.06  & -0.64 $\pm$   8   \\
    &   6   &   19.7  &  57515.45 &   19.74 &   69960.89&   22.4$\pm$   8.4 \\
    &   7   &   21.37 &   30765.45&   21.36 &   54300.82&  21.89$\pm$   8   \\
    &   8   &   22.55 &   23176.57&   22.92 &   41652.65&   0.8 $\pm$   8   \\
    &   9   &   23.38 &   27241.76&   24.1  &   61314.25& -12.2 $\pm$   14.6   \\
    &   10  &   24.65 &   26648.13&   24.58 &   60490.65& -13.54$\pm$   8   \\
    &   11  &   26.34 &   35474.09&   26.34 &   83907.04&   9.54$\pm$   8   \\
    &   12  &   28.06 &   23596.59&   28.19 &   51085.72& -22.46$\pm$   8   \\
    &   13  &   29.22 &   28968.98&   29.22 &   62544.48&   8.47$\pm$   8   \\
    &   14  &   31.11 &   37284.11&   33.8  &   57153.23&   5.8 $\pm$   8    \\
    &   15  &   35.38 &   26963.79&   35.98 &   59132.34&   25.0$\pm$   8   \\  \hline
7185&   1   &   5.49  &   6175.93 &   5.78  &   5385.61 & 100.8 $\pm$   8   \\
    &   2   &   53.18 &   6798.85 &   52.46 &   6463    & 182.4 $\pm$   8   \\
    &   3   &   67.7  &   7703.87 &   67.68 &   6414    & 294.53$\pm$   8   \\
    &   4   &   69.89 &   7398.18 &   69.75 &   8126.25 &  252.8$\pm$   8   \\
    &   5   &   74.99 &   6745.86 &   74.73 &   5792.94 & 190.08$\pm$   8.45    \\
    &   6   &   86.06 &   7479.47 &   85.94 &   5315.92 & 228.8 $\pm$   8   \\
    &   7   &   90.67 &   5665.2  &   90.7  &   4376.22 & 224.35$\pm$   8   \\
    &   8   & 107.66  &   6722.01 & 107.16  &   4076.01 & 386.4 $\pm$   11.2    \\
    &   9   &  110.09 &   7238.85 &   109.63&   4458.83 & 276.67$\pm$   12.58   \\
    &   10  & 131.69  &   5886.4  &   131.31&   4664.4  & 476.8 $\pm$   32  \\
    &   11  & 134.63  &   7530.29 & 134.25  &   4752.98 & 630.72$\pm$   14.02   \\
    &   12  & 141.51  &   6691.83 &   140.93&   4029.32 &   456 $\pm$   12  \\
    &   13  &   148.04&   10601.5 &   147.58&   8035.55 & 441.6 $\pm$   8   \\
    &   14  &   155.62&   6341.69 &   154.6 &   3846.78 & 427.58$\pm$   12.58   \\  \hline
7240&   1   &   0.66  &   6143.53 &   0.62  &   15187.18&   5.85$\pm$   8   \\
    &   2   &   3.02  &   9470.03 &   3.02  &   22659.87& 39.04 $\pm$   8   \\  \hline
7247&   1   &   13.83 &   4890.78 &   20.24 &   6322.53 & -409.6$\pm$   64  \\
    &   2   &   35.34 &   6156.84 &   35.2  &   10162.96& 140.8 $\pm$   9.6 \\  \hline
7277&   1   &   5.81  &   8330.56 &   5.62  &   7526.75 & 78.62 $\pm$   11.23   \\
    &   2   &   12.06 &   9470.15 &   11.91 &   8152.23 & 35.33 $\pm$   11.78   \\
    &   3   &   13.14 &  10034.39 &   13.11 &   8292.93 & 16.93 $\pm$   8.47    \\
    &   4   &   13.93 &   9891.21 &   13.94 &   8363.95 & -28.16$\pm$   24.64   \\
    &   5   &   15.56 &   9805.23 &   15.56 &   10040.36&   -28 $\pm$   8.4 \\
    &   6   &   16.94 &   10139.89&   16.9  &   10283.65& -8.38 $\pm$   8   \\
    &   7   &   23.52 &   10585.91&   23.5  &   9677.3  &   11.9$\pm$   11.9    \\
    &   8   &   24.32 &  10334.18 &   24.16 &   9153.41 & -10.09$\pm$   8   \\
    &   9   &   25.78 &  14267.59 &   25.73 &   12911.01&-15.74 $\pm$   11.81   \\
    &   10  &   27.94 & 10905.58  &   27.38 &   10817.12& -92.8 $\pm$   32   \\  \hline
7301&   1   &   2.08  & 13211.36  &   3.3   &   15796.25&   33.5$\pm$   8.1   \\
    &   2   &   16.9  & 31436.79  &   16.67 &   90769.16&   10.7$\pm$   8   \\
    &   3   &   19.08 &   28238.96&   19.09 &   79843.72& -21.9 $\pm$   8   \\
    &   4   &   32.45 &   13325.43&   32.57 &   12153.23&-113.6 $\pm$   8.4 \\  \hline
7343&   1   &   26.23 &   13877.48&   25.34 &   35294.72&   33.6$\pm$   16.8    \\
    &   2   &   32.1  &   9970.74 &   31.39 &   18137.39& 235.87$\pm$   8   \\
    &   3   &   38.73 & 11990.98  &   38.3  &   29087.97& 313.82$\pm$   22.42   \\
    &   4   &   77.9  &   9681.45 &   76.88 &   11955.15&   88  $\pm$   36  \\  \hline
7560&   1   &   4.63  &   13276.4 &   4.83  &   14945.85& -5.11 $\pm$   8 \\
    &   2   &   6.61  & 10286.24  &   6.47  &   6738.54 & -21.9 $\pm$   12.2   \\
    &   3   &   45.45 &   14452.45&   45.22 &   19496.01&   160 $\pm$   8   \\
    &   4   &   48.34 &   15433.98&   47.94 &   20457.11& 159.3 $\pm$   8   \\
    &   5   &   58.28 &   10509.12&   58.25 &   6044.53 & 226.37$\pm$   8   \\
    &   6   &   72.34 &   7915.58 &   71.73 &   5624.59 &181.38 $\pm$   17.44   \\  \hline
7281&   1   &   0.18  &   19313.04&   0.08  &   35588.5 &56.99  $\pm$   56.45   \\
    &   2   &   0.56  &   24548.62&   0.56  &   29636.68& 17.92 $\pm$   8   \\
    &   3   &   1.38  &   14841.34&   1.39  &   11134.64&  -5.71$\pm$   8   \\  \hline
7592&   1   &   1.1   &   28885.81&   1.1   &   87186.68&   12.8$\pm$   8   \\
    &   2   &   2.42  &   21719.17&   2.42  &   52149.73&   25.6$\pm$   8   \\
    &   3   &   25.54 &   19134.16&   24.9  &   20753.14&   568 $\pm$   40.8    \\  \hline
7605&   1   &   1     &   6280.25 &   0.82  &   6372.43 &   47.2$\pm$   9.6 \\
    &   2   &   2.06  &   7473.61 &   2.04  &   5608.95 & 79.03 $\pm$   8   \\
    &   3   &   2.84  &   6769.1  &   2.56  &   7014.72 &109.66 $\pm$   8   \\
    &   4   &   5.09  &   6219.77 &   7.33  &   5920.6  &   65.0$\pm$   15.8    \\
    &   5   &   10.47 &   7038.2  &   10.45 &   6106.67 & 25.34 $\pm$   8.45    \\
    &   6   &   12.55 &   6421.32 &   12.49 &   6028.94 & 67.07 $\pm$   8   \\
    &   7   &   13.91 &   7052.97 &   14.35 &   4542.31 & -3.87 $\pm$   33.9   \\
    &   8   &   16.1  &   6217.25 &   16.07 &   6229.82 &  68.08$\pm$   8   \\
    &   9   &   16.67 &   6533.59 &   16.86 &   6651.85 & 39.68 $\pm$   11.9    \\  \hline
7610&   1   &   2.63  &   7641.71 &   2.54  &   8291.97 &  67.74$\pm$   8   \\
    &   2   &   3.65  &   9039.63 &   3.79  &   10293.75&  40.32$\pm$   8   \\
    &   3   &   4.76  &   10056.44&   4.7   &   21288.27&   3.2 $\pm$   8   \\
    &   4   &   5.46  &   5542.52 &   5.45  &   9148.14 & 148.72$\pm$   22.12   \\  \hline
7651&   1   &   2.63  &   6889.28 &   2.44  &   8383.71 & 130.48$\pm$   13.5    \\
    &   2   &   3.98  &   7423.85 &   3.55  &   10129.12& 12.48 $\pm$   21.6    \\
    &   3   &   7.73  &   8513.08 &   7.49  &   8494.76 & 118.88$\pm$   8   \\
    &   4   &   27.58 &   7887.49 &   27.42 &   6972.15 & 202.24$\pm$   23.04   \\
    &   5   &   28.26 &   9172.29 &   28.21 &   11805.66& 98.78 $\pm$   8   \\
    &   6   &   30.02 &   9383.27 &   29.71 &   9821.42 & 265.57$\pm$   15.44   \\
    &   7   &   31.97 &   8674.51 &   31.66 &   9437.04 & 162.4 $\pm$   8.4 \\
    &   8   &   34.05 &   9159.45 &   33.9  &   12863.55& 134.4 $\pm$   8.4 \\  \hline
7688&   1   &   3.14  &   5021.42 &   4.09  &   6534.36 &  67.36$\pm$   12.96   \\
    &   2   &   7.16  &   5773.21 &   6.75  &   10716.66&  74.24$\pm$   10.24   \\
    &   3   &   11.48 &   4919.93 &   11.02 &   7405.73 & 99.84 $\pm$   15.36   \\
    &   4   &   18.82 &   4765.4  &   18.62 &   6767.25 &  68.8 $\pm$   8   \\  \hline
7695&   1   &   0.01  &   9997.59 &   0.14  & 34082.12  &  7.68 $\pm$   8   \\
    &   2   &   1.26  &   15184.9 &   1.26  & 47648.71  &   9.6 $\pm$   8   \\
    &   3   &   2.08  &   11189.21&   1.93  & 28174.99  & -0.93 $\pm$   8   \\
    &   4   &   2.26  &  12987.99 &   2.26  & 41957.11  &  6.38 $\pm$   8   \\
    &   5   &   4.14  &   15044.82&   2.82  & 41102.16  &  -2.82$\pm$   8   \\
    &   6   &   5.94  &   13755.44&   5.94  & 39713.14  & 1.28  $\pm$   8   \\
    &   7   &   7.02  &   12637.03&   7.09  & 26390.67  & -6.4  $\pm$   8   \\
    &   8  &   9.15  &   8964.57 &   8.94  & 13251.85  & 2.56  $\pm$   8   \\
    &   9  &   9.9   &   8043.57 &   9.81  & 10469.08  & -14.08$\pm$   8   \\
    &   10  &   11.7  &   10652.57&   11.76 &   18560.27& -10.88$\pm$   8   \\
    &   11  &   13.89 &   8558.67 &   13.91 &   10525.66& -5.12 $\pm$   8   \\
    &   12  &   14.67 &   7621.24 &   14.46 &   11840.41&  9.8  $\pm$   8   \\
    &   13  &   21.69 &   7419.13 &   21.81 &   8900.46 & -26.88$\pm$   12.48   \\  \hline
7906&   1   &   17.26 &   27293.02&   17.13 &   52387.8 & 154.18$\pm$   8   \\
    &   2   &   20.58 &   62602.4 &   20.52 & 130010.53 & 92.22 $\pm$   8.38    \\
    &   3   &   22.71 &   62967.53&   22.62 & 125032.32 & 78.72 $\pm$   8   \\
    &   4   &   25.72 &   29331.29&   25.64 &   30459.1 &   140 $\pm$   8   \\
    &   5   &   26.82 &   28886.48&   26.78 &   19976.09& 89.38 $\pm$   8   \\
    &   6   &   29.68 &   49836.62&   29.64 &   69791.91&  59.36$\pm$   8   \\
    &   7   &   31.3  &   51076.85&   31.25 &   63184.7 &  93.44$\pm$   8   \\  \hline
7925&   1   &   9.99  &   8833.15 &   9.5   &  16207.43 & 192.0 $\pm$   38  \\
    &   2   &   28.32 &   6167.3  &   28.29 &   11008.93& 81.85 $\pm$   8   \\
    &   3   &   30.42 &   6465.81 &   30.48 &   7932.42 &   13.6$\pm$   12.32   \\
    &   4   &   31.15 &   6636.39 &   30.89 &   7807.2  &  104.4$\pm$   14.11   \\
    &   5   &   87    &   7829.75 &   86.98 &  14396.44 & 441.6 $\pm$   12.8    \\  \hline
7954&   1   &   0.85  &   18555.32&   0.8   &  36346.06 &   48  $\pm$   8   \\
    &   2   &   1.23  &   15929.81&   1.24  &   53627.18&13.76  $\pm$   8   \\
    &   3   &   9.45  &   30656.88&   9.42  &   68396.69&  60.48$\pm$   8   \\
    &   4   &   10.14 &   22981.44&   10.12 &   53973.07&  41.92$\pm$   8   \\
    &   5   &   10.73 &   31529.55&   10.73 &   86100.83& 37.76 $\pm$   8   \\
    &   6   &   12    &   10284.52&   12    &   16411.95&  -2.25$\pm$   8   \\
    &   7   &   12.85 &   21002.68&   12.84 &   32831.5 &   22.5$\pm$   8   \\
    &   8   &   13.89 &   18092.25&   13.89 &   34841.63&  47.24$\pm$   8   \\
    &   9   &   14.87 &   18788.42&   14.81 &   45039.05&  24.75$\pm$   8   \\  \hline
7975&   1   &   5.62  &   5948.9  &   4.24  &   5836.87 & 268.8 $\pm$   43.2    \\
    &   2   &   69.78 &   5117.93 &   69.9  &   5512.82 & 153.6 $\pm$   8  \\
    &   3   &   75.06 &   4831.78 &   74.44 &   4069.48 &284.26 $\pm$   15.79   \\
    &   4   &   77.93 &   8232.47 &   77.89 &   11267.32&191.23 $\pm$   10.62   \\
    &   5   &   78.84 &   8207.79 &   79.46 &   9780.49 & 168.2 $\pm$   54.7   \\
    &   6   &   80.38 &   10760.89&   80.14 &   15902.9 & 42.08 $\pm$   8   \\
    &   7   &   81.38 &   10250.06&   81.14 &   13218   &154.56 $\pm$   8   \\
    &   8   &   83.45 &   10295.42&   83.31 &   10273.4 & 252.1 $\pm$   12.93   \\
    &   9   &   89.46 &   6073.21 &   89    &   5618.15 & 212.8 $\pm$   8   \\
    &   10  &   92.61 &   6771.17 &   92.41 &   6023.46 & 218.4 $\pm$   8   \\
    &   11  &   95.56 &   7030.29 &   95.44 &   6191.42 & 231.84$\pm$   13.44   \\
    &   12  &   97.02 &   8516.03 &   96.9  &   7297    & 212.0 $\pm$   8   \\  \hline
8022&   1   &   2.83  &   8080.59 &   2.78  &   11553.08& 107.84$\pm$   30.24   \\
    &   2   &   3.38  &   6685.64 &   3.29  &   5216.47 & -44.8 $\pm$   39.2    \\
    &   3   &   7.82  &   7066.22 &   7.36  &   4570.54 & 117.38$\pm$   8.38    \\
    &   4   &   13.25 &   6304.86 &   13.74 &   4473.76 & 97.66 $\pm$   17.44   \\  \hline
8030&   1   &   1.02  &   4732.8  &   0.95  &   7478.56 & 66.82 $\pm$   8.35    \\
    &   2   &   2.65  &   4164.49 &   2.63  &   3862.87 & 122.5 $\pm$   8.45    \\
    &   3   &   15.6  &   4580.29 &   15.62 &   7527.69 & 161.28$\pm$   8   \\
    &   4   &   17.51 &   4492.37 &   17.14 &   6397.1  & 43.68 $\pm$   8   \\
    &   5   &   18.18 &   4500.05 &   18.14 &   4849.06 & 43.3  $\pm$   11.81   \\
    &   6   &   20.3  &   4983.15 &   20.31 &   10017.68& 117.38$\pm$   8   \\  \hline
8081&   1   &   0.75  &   5776.88 &   0.71  &   6822.93 & 27.84 $\pm$   8   \\
    &   2   &   20.7  &   8202.56 &   20.75 &   10101.55& 71.81 $\pm$   8   \\
    &   3   &   21.94 &   8494.25 &   21.9  &   11999.75& 25.31 $\pm$   8   \\
    &   4   &   23.34 &   7230.99 &   23.26 &   5887.67 & 107.52$\pm$   8   \\  \hline
8087&   1   &   1.2   &   5204.05 &   1.3   &   5597.91 &   76  $\pm$   34  \\
    &   2   &   5.12  &   7713.57 &   5.1   &   11056.65&   64  $\pm$   8   \\
    &   3   &   6.9   &   6259.22 &   6.5   &   8891.9  & 139.87$\pm$ 8   \\
    &   4   &   8.86  &   5572.11 &   8.72  &   8376.39 &   0   $\pm$ 8   \\
    &   5   &   10.1  &   5753.64 &   10.03 &   7333.64 &   13.6$\pm$ 16.72\\
    &   6   &   40    &   5275.8  &   39.66 &   4921.9  &   26.6$\pm$ 11.4 \\  \hline
8109&   1   &   0.57  &   6507.3  &   0.23  &   7823.73 & 227.2 $\pm$ 8   \\
    &   2   &   2.36  &   7943.62 &   1.26  &   10602.3 & 824.84$\pm$ 68.2\\  \hline
\enddata
\tablecomments{From the left to right columns:\\
Column 1: the GRB trigger number tabulated in the `{\it BATSE} Catalog';\\
Column 2: the numeral identification of
an individual pulse in each GRB event;\\
Columns 3 \& 5: the pulse peak positions of the low-energy band
($25-55$ keV) and of the high-energy band ($110-320$ keV), respectively;\\
Columns 4 \& 6: the peak count rates of the low-energy band and of
the high-energy band, respectively, where the peak count rate is
determined by the average of three bins around the maximum;\\
Column 7: the spectral lags with their error estimates.}
\end{deluxetable}

\end{document}